\documentclass[a4paper,10pt]{article}

\usepackage[latin1]{inputenc}
\usepackage{ucs}
\usepackage[T1]{fontenc}
\usepackage{bm}
\usepackage{upgreek}
\usepackage{color}
\usepackage{mathrsfs} 
\usepackage{pgfplots}
\usepackage{amsmath}
\usepackage{amsfonts}
\usepackage{amssymb}
\usepackage{graphicx}
\usepackage{Formulas}
\usepackage[round,comma]{natbib}
\usepackage{pdfsync}
\usepackage{srcltx}

\newcommand\vm[1]{\bm{\mathrm{#1}}} 
\newcommand{\red}[1]{\textcolor{black}{#1}} 

\begin{document}

\begin{center}
 \Large{On the role of enrichment and statical admissibility of recovered fields in a-posteriori error estimation for enriched finite element methods}
\end{center}

\begin{center}
 Octavio A. González-Estrada$^{1}$, Juan~Jos\'{e} R\'{o}denas$^{2}$, St\'{e}phane P.A. Bordas$^{1}$, Marc~Duflot$^{3}$, Pierre Kerfriden$^{1}$, Eugenio Giner$^{2}$
\end{center}

\begin{center}
\begin{small}
$^{1}$Institute of Mechanics and Advanced Materials. Cardiff School of Engineering, Cardiff University, The Parade, Cardiff CF24 3AA Wales, UK.\\ 
$^{2}$Centro de Investigación de Tecnología de Vehículos(CITV),
\\Universitad Politècnica de València, E-46022-Valencia, Spain.\\
$^{3}$CENAERO, Rue des Frères Wright 29, B-6041 Gosselies, Belgium
\end{small}
\end{center}

\begin{abstract}
\noindent \textbf{Purpose} -- This paper aims at assessing the effect of (1) the statical admissibility of the recovered solution; (2) the ability of the recovered solution to represent the singular solution; on the accuracy, local and global effectivity of recovery-based error estimators for enriched finite element methods (e.g. the extended finite element method, XFEM). \\
\textbf{Design/methodology/approach} -- We study the performance of two recovery techniques. The first is a recently developed superconvergent patch recovery procedure with equilibration and enrichment (SPR-CX). The second is known as the extended moving least squares recovery (XMLS), which enriches the recovered solutions but does not enforce equilibrium constraints. Both are extended recovery techniques as the polynomial basis used in the recovery process is enriched with singular terms for a better description of the singular nature of the solution.\\ 
\textbf{Findings} -- Numerical results comparing the convergence and the effectivity index of both techniques with those obtained without the enrichment enhancement clearly show the need for the use of extended recovery techniques in Zienkiewicz-Zhu type error estimators for this class of problems. The results also reveal significant improvements in the effectivities yielded by statically admissible recovered solutions.\\
\textbf{Originality/value} --  This work shows that both extended recovery procedures and statical admissibility are key to an accurate assessment of the quality of enriched finite element approximations. 
\end{abstract}

{\textbf{Paper Type} Research Paper}

\section{Introduction}

Engineering structures, in particular in aerospace engineering, are intended to operate with flawless components, especially for safety critical parts. However, there is always a possibility that cracking will occur during operation, risking catastrophic failure and associated casualties.

The mission of Damage Tolerance Assessment (DTA) is to assess the influence of these defects, cracks and damage on the ability of a structure to perform safely and reliably during its service life. An important goal of DTA is to estimate the fatigue life of a structure, i.e. the time during which it  remains safe given a pre-existing flaw.

Damage tolerance assessment relies on the ability to accurately predict crack paths and growth rates in complex structures. Since the simulation of three-dimensional crack growth is either not supported by commercial software, or requires  significant effort and time for the analysts  and is generally not coupled to robust error indicators, reliable, quality-controlled, industrial damage tolerance assessment is still a major challenge in engineering practice. 

The extended finite element method (XFEM) \citep{Moes1999} is now one of many successful numerical methods to solve fracture mechanics problems. The particular advantage of the XFEM, which relies on the partition of unity (PU) property \citep{Melenk1996} of finite element shape functions, is its ability to model cracks without the mesh conforming to their geometry. This allows the crack to split the background mesh arbitrarily, thereby leading to significantly increased freedom in simulating crack growth. 

This feat is achieved by adding new degrees of freedom to (i) describe the discontinuity of the displacement field across the crack faces, within a given element, and (ii) reproduce the asymptotic fields around the crack tip. Thanks to the advances made in the XFEM during the last years, the method is now considered to be a robust and accurate means of analysing fracture problems, has been implemented in commercial codes \citep{abaqus2009,cenaero2011} and is industrially in use for damage tolerance assessment of complex three dimensional structures \citep{Bordas2006,Wyart2007,Duflot2007}.

While XFEM allows to model cracks without (re)meshing the crack faces as they evolve and yields exceptionally accurate stress intensity factors (SIF) for 2D problems,  \cite{Bordas2006,Wyart2007,Duflot2007} show  that for realistic 3D structures, a ``very fine'' mesh is required to accurately capture the complex three-dimensional stress field and  obtain satisfactory stress intensity factors (SIFs), the main drivers of linear elastic crack propagation. 

Consequently, based on the mesh used for the stress analysis, a new mesh offering sufficient refinement throughout the whole potential path of the crack must be constructed \emph{a priori},  i.e. before the crack path is known. Practically, this is done by running  preliminary analyses on  coarse meshes to obtain an approximative crack path and heuristically refining the mesh around this path to increase accuracy.
 
Typically, this refinement does not rely on sound error measures, thus the heuristically chosen mesh is in general inadequately suited and can cause large inaccuracies in the crack growth path, especially around holes, which can lead to non-conservative estimates of the safe life of the structure. 

Thus, it is clear that although XFEM simplifies the treatment of cracks compared to the standard FEM by lifting the burden of a geometry conforming mesh, it  still requires iterations and associated user intervention and it employs heuristics which are detrimental to the robustness and accuracy of the simulation process.

It would be desirable to minimise the changes to the  mesh topology, thus user intervention, while ensuring the stress fields and SIFs are accurately computed at each crack growth step. This paper is one more step in attempting to control the discretization error committed by enriched FE approximations,  decrease human intervention in damage tolerance assessment of complex industrial structures and enhance confidence in the results by providing enriched FEMs with sound error estimators which will guarantee a predetermined accuracy level and suppress recourse to manual iterations and heuristics.

The error assessment procedures used in finite element analysis are well known and can be usually classified into different families \citep{Ainsworth2000}: residual based error estimators, recovery based error estimators, dual techniques, \textit{etc}. Residual based  error estimators were substantially improved with the introduction of the  \emph{residual equilibration} by \cite{Ainsworth2000}. These error estimators have a strong mathematical basis and have been frequently used to obtain lower and upper bounds of the error \citep{Ainsworth2000,Diez2004}. Recovery based error estimates were first introduced by \cite{Zienkiewicz1987} and are often preferred by practitioners because they are robust and simple to use and provide an enhanced solution. Further improvements were made with the introduction of new recovery processes such as the superconvergent patch recovery (SPR) technique proposed by \cite{Zienkiewicz1992, Zienkiewicz1992b}. Dual techniques based on the evaluation of two different fields, one compatible in displacements and another equilibrated in stresses, have also been used by \cite{Pereira1999} to obtain bounds of the error. Herein we are going to focus on recovery based techniques which follow the ideas of the Zienkiewicz-Zhu (ZZ) error estimator proposed by \cite{Zienkiewicz1987}.

The literature on error estimation techniques for mesh based partition of unity methods is still scarce. One of the first steps in that direction was made in the context of the Generalized Finite Element Method (GFEM) by \cite{Strouboulis2001}. The authors proposed a recovery-based error estimator  which provides good results for \textit{h}-adapted meshes. In a later work two new \textit{a posteriori} error estimators for GFEM were presented \citep{Strouboulis2006}. The first one was based on patch residual indicators and provided an accurate theoretical upper bound estimate, but its computed version severely underestimated the exact error. The second one was an error estimator based on a recovered displacement field and its performance was closely related to the quality of the GFEM solution. This recovery technique constructs a recovered solution on patches using a basis enriched with handbook functions.

In order to obtain a recovered stress field that improves the accuracy of the stresses obtained by XFEM, \cite{Xiao2006} proposed a moving least squares fitting adapted to the XFEM framework which considers the use of statically admissible basis functions. Nevertheless, the recovered stress field was not used to obtain an error indicator.

\cite{Pannachet2008} worked on error estimation for mesh adaptivity in XFEM for cohesive crack problems. Two error estimates were used, one based on the error in the energy norm and another that considers the error in a local quantity of interest. The error estimation was based on solving a series of local problems with prescribed homogeneous boundary conditions.

\cite{Panetier2010} presented an extension to enriched approximations of the constitutive relation error (CRE) technique already available to evaluate error bounds in FEM \red{\citep{ladevezepelle2005}}. This procedure has been used to obtain local error bounds on quantities of interest for XFEM problems.

\cite{Bordas2007} and \cite{Bordas2008} proposed a recovery based error estimator for 2D and 3D XFEM approximations for cracks known as the extended moving least squares (XMLS). This method intrinsically enriched an MLS formulation to include information about the singular fields near the crack tip. Additionally, it used a diffraction method to introduce the discontinuity in the recovered field. This error estimator provided accurate results with effectivity indices close to unity (optimal value) for 2D and 3D fracture mechanics problems. Later,  \cite{Duflot2008} proposed a global derivative recovery formulation extended to XFEM problems. The recovered solution was sought in a space spanned by the near tip strain fields obtained from differentiating the Westergaard asymptotic expansion. Although the results provided by this technique were not as accurate as those in \cite{Bordas2007, Bordas2008}, they were deemed by the authors to require less computational power. 

\cite{Rodenas2007a} presented a modification of the superconvergent patch recovery (SPR) technique tailored to the XFEM framework called $\textrm{SPR}_{\textrm{XFEM}}$. This technique was based on three key ingredients: (1) the use of a \emph{singular}+\emph{smooth} stress field decomposition procedure around the crack tip similar to that described by \cite{Rodenas2006} for FEM; (2) direct calculation of recovered stresses at integration points using the partition of unity, and (3) use of different stress interpolation polynomials at each side of the crack when a crack intersects a patch. In order to obtain an equilibrated smooth recovered stress field, a simplified version of the SPR-C technique presented by \cite{Rodenas2007} was used. This simplified SPR-C imposed the fulfilment of the boundary equilibrium equation at boundary nodes but did not impose the satisfaction of the internal equilibrium equations. 

The numerical results presented by \cite{Bordas2007, Bordas2008} and \cite{Rodenas2007a} showed promising accuracy for both the XMLS and the $\textrm{SPR}_{\textrm{XFEM}}$ techniques. It is  apparent in those papers that  $\textrm{SPR}_{\textrm{XFEM}}$ led to effectivity indices remarkably close to unity. Yet, these papers do not shed any significant light on the respective roles played by two important ingredients in those error estimators, namely:
\begin{enumerate}
\item The statistical admissibility of the recovered solution;
\item The enrichment of the recovered solution.
\end{enumerate} 

Moreover, the differences in the test cases analysed and in the quality measures considered in each of the papers makes it difficult to objectively compare the merits of both methods.  

The aim of this paper is to assess the role of statistical admissibility and enrichment of the recovered solution in recovery based error estimation of enriched finite element approximation for linear elastic fracture. To do so, we perform a systematic study of the results obtained when considering the different features of two error estimation techniques: XMLS and SPR-CX. SPR-CX is an enhanced version of the $\textrm{SPR}_{\textrm{XFEM}}$  presented by \cite{Rodenas2007a} for 2D problems. Advantages and disadvantages of each method are also provided. 

The outline of the paper is as follows. In Section 2, the XFEM is briefly presented. Section 3 deals with error estimation and quality assessment of the solution.  In Sections 4 and 5 we introduce the error estimators used in XFEM approximations: the SPR-CX and the XMLS, respectively. Some numerical examples analysing both techniques and the effect of the enrichment functions in the recovery process are presented in Section 6. Finally some concluding remarks are provided in Section 7.

\section{Reference problem and XFEM solution}

Let us consider a 2D linear elastic fracture mechanics (LEFM) problem on a bounded domain $\Omega \subset \mathbb{R}^2$. The unknown displacement field $\vm{u}$ is the solution of the boundary value problem 

\begin{align}
	\nabla \cdot \boldsymbol{\upsigma} (\vm{u})+\vm{b} &= \mathbf{0} &\quad&\textrm{in } \Omega\label{Eq:IntEq}\\
	 \boldsymbol{\upsigma}(\vm{u})\cdot\vm{n} &= \vm{t}       &&\textrm{on } \Gamma_N \label{Eq:Neumann}\\
	 \boldsymbol{\upsigma}(\vm{u})\cdot\vm{n} &= \mathbf{0}         &&\textrm{on } \Gamma_C \label{Eq:NeumannCrack}\\
	\vm{u} &= \bar{\vm{u}}        &&          \textrm{on } \Gamma_D \label{Eq:Dirichlet}
\end{align}

\noindent where $\Gamma_N$ and $\Gamma_D$ are the Neumann and Dirichlet boundaries and $\Gamma_C$ represents the crack faces, with $\partial \Omega = \Gamma_N \cup \Gamma_D  \cup \Gamma_C$ and $\Gamma_N \cap \Gamma_D  \cap \Gamma_C=\varnothing$. $\vm{b}$ are the body forces per unit volume, $\vm{t}$ the tractions applied on $\Gamma_N$ (being $\vm{n}$ the normal vector to the boundary) and $\bar{\vm{u}}$ the prescribed displacements on $\Gamma_D$. The weak form of the problem reads: Find $\vm{u} \in V$ such that:
\begin{equation}
\forall\vm{v} \in V \qquad a(\vm{u},\vm{v})=l(\vm{v}), 
\end{equation}

\noindent where $V$ is the standard test space for elasticity problems such that \red{$V = \{\vm{v} \;|\; \vm{v} \in  H^1(\Omega) , \vm{v}|_{\Gamma_D}(\vm{x}) = \mathbf{0}, \vm{v}\;\textrm{discontinuous on } \Gamma_C \}$}, and 

\begin{align}
a(\vm{u},\vm{v})  & := \int _{\Omega}  \boldsymbol{\upsigma} (\vm{u}) : \vm{\epsilon}(\vm{v}) d \Omega = 
\int _{\Omega}  \boldsymbol{\upsigma}(\vm{u})  : \boldsymbol{\mathsf{S}} :  \boldsymbol{\upsigma}(\vm{v}) d \Omega \\
l(\vm{v})&:=\int _{\Omega} \vm{b} \cdot \vm{v}d \Omega + \int _{\Gamma_N} \vm{t} \cdot \vm{v}d \Gamma,
\end{align}

\noindent where $\boldsymbol{\mathsf{S}}$ is the compliance tensor, $ \boldsymbol{\upsigma}$ and $\vm{\epsilon}$ represent the stress and strain operators.

LEFM problems are denoted by the singularity present at the crack tip. The following expressions represent the first term of the asymptotic expansion which describes the displacements and stresses for combined loading modes I and II in 2D. These expressions can be found in the literature \citep{Szabo1991, Rodenas2007a} and are reproduced here for completeness:

\begin{equation}\label{Eq:CrackTipField:Disp}
\begin{aligned}
u_1(r,\phi)&=\frac{K_{\rm I}}{2\mu} \sqrt{\frac{r}{2\pi }} \cos\frac{\phi }{2} \left(\kappa-\cos \phi\right)+\frac{K_{\rm II}}{2\mu} \sqrt{\frac{r}{2\pi}} \sin\frac{\phi}{2}\left(2+\kappa+ \cos \phi \right) \\
u_2(r,\phi)&=\frac{K_{\rm I}}{2\mu} \sqrt{\frac{r}{2\pi }} \sin\frac{\phi }{2} \left(\kappa-\cos \phi\right)+\frac{K_{\rm II}}{2\mu} \sqrt{\frac{r}{2\pi}} \cos\frac{\phi}{2}\left(2-\kappa- \cos \phi \right)
\end{aligned}
\end{equation}

\begin{equation}\label{Eq:CrackTipField:Stress}
\begin{aligned}
\sigma_{11}(r,\phi)&=\frac{K_{\rm I}}{\sqrt{2\pi r}}\cos\frac{\phi}{2}\left(1-\sin\frac{\phi}{2} \sin\frac{3\phi}{2}\right)-\frac{K_{\rm II}}{\sqrt{2\pi r}}\sin\frac{\phi}{2}\left( 2+\cos\frac{\phi}{2}\cos \frac{3\phi}{2} \right)\\
\sigma_{22}(r,\phi)&=\frac{K_{\rm I}}{\sqrt{2\pi r}} \cos\frac{\phi}{2} \left(1+\sin\frac{\phi}{2} \sin\frac{3\phi}{2}\right)+\frac{K_{\rm II}}{\sqrt{2\pi r}}\sin\frac{\phi}{2} \cos\frac{\phi}{2}\cos \frac{3\phi}{2}\\
\sigma_{12}(r,\phi)&=\frac{K_{\rm I}}{\sqrt{2\pi r}} \sin\frac{\phi}{2} \cos\frac{\phi}{2} \cos \frac{3\phi}{2} + \frac{K_{\rm II}}{\sqrt{2\pi r}}\cos\frac{\phi}{2} \left(1-\sin\frac{\phi}{2} \sin\frac{3\phi}{2}\right)
\end{aligned}
\end{equation}

\noindent where $r$ and $\phi$ are the crack tip polar coordinates, $K_{\rm I}$ and $K_{\rm II}$ are the stress intensity factors for modes I and II, $\mu$ is the shear modulus, and  $\kappa$ the Kolosov's constant, defined in terms of the parameters of material $E$ (Young's modulus) and $\upsilon$ (Poisson's ratio), according to the expressions:

\begin{equation*}
\mu =\frac{E}{2\left(1+\upsilon \right)}, \qquad 
\kappa=\left\{\begin{array}{l c r} {\displaystyle 3-4\upsilon \qquad \textrm{plane strain}}\\ \noalign{\medskip}{\displaystyle \frac{3-\upsilon}{1+\upsilon}\, \qquad \textrm{plane stress}} \end{array}\right.
\end{equation*}

\begin{figure}[!ht]
	\centering
		 \includegraphics{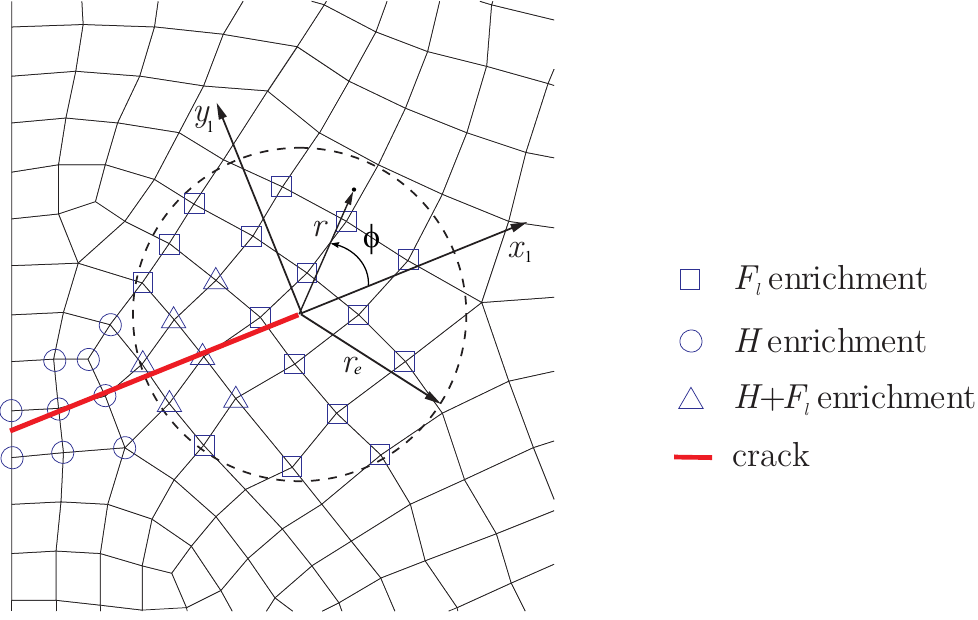} 
	\caption{Classification of nodes in XFEM. Fixed enrichment area of radius $r_e$}
	\label{fig:XFEM_enrichment}
\end{figure}

This type of problem is difficult to model using a standard FEM approximation as the mesh needs to explicitly conform to the crack geometry. With the XFEM the discontinuity of the displacement field along the crack faces is introduced by adding degrees of freedom to the nodes of the elements intersected by the crack. This tackles the problem of adjusting the mesh to the geometry of the crack \citep{Moes1999, Stolarska2001}. Additionally, to describe the solution around the crack tip, the numerical model introduces a basis that spans the near tip asymptotic fields. The following expression is generally used to interpolate the displacements at a point of coordinates $\vm{x}$ accounting for the presence of a crack tip in a 2D XFEM approximation:

\begin{equation} \label{Eq:u-XFEM} 
\vm{u}_{h}(\vm{x}) =\sum _{i\in  \mathcal{I}}N_{i}(\vm{x}) \textbf{a}_{i}  +\sum _{j\in  \mathcal{J}}N_{j}(\vm{x}) H(\vm{x})\textbf{b}_{j}  +\sum _{m\in  \mathcal{M}}N_{m}(\vm{x}) \left(\sum _{\ell =1}^{4}F_{\ell } (\vm{x})\textbf{c}_{m}^{\ell }  \right)  
\end{equation}

\noindent where $N_i$ are the shape functions associated with node $i$, $\textbf{a}_{i}$ represent the conventional nodal degrees of freedom, $\textbf{b}_{j}$ are the coefficients associated with the discontinuous enrichment functions, and $\vm{c}_{m}$ those associated with the functions spanning the asymptotic field. In the above equation, $ \mathcal{I}$ is the set of all the nodes in the mesh, $ \mathcal{M}$ is the subset of nodes enriched with crack tip functions, and $ \mathcal{J}$ is the subset of nodes enriched with the discontinuous enrichment (see Figure~\ref{fig:XFEM_enrichment}). In \eqref{Eq:u-XFEM}, the Heaviside function $H$, with unitary modulus and a change of sign on the crack face, describes the displacement discontinuity if the finite element is intersected by the crack. The $F_{\ell }$ are the set of branch functions used to represent the asymptotic expansion of the displacement field around the crack tip seen in (\ref{Eq:CrackTipField:Disp}). For the 2D case, the following functions are used \citep{Belytschko1999}:

\begin{equation} \label{Eq:branch_functions} 
\left\{F_{\ell } \left(r,\phi \right)\right\}\equiv \sqrt{r} \left\{\sin\frac{\phi }{2} ,\cos \frac{\phi }{2} ,\sin\frac{\phi }{2} \sin\phi ,\cos \frac{\phi }{2} \sin\phi \right\} 
\end{equation} 

The main features of the XFEM implementation considered to evaluate the numerical results is described in detail in \cite{Rodenas2007a} and can be summarized as follows:

\begin{itemize}
\item Use of bilinear quadrilaterals.
\item Decomposition of elements intersected by the crack into integration subdomains that do not contain the crack. Alternatives which do not required this subdivision are proposed by \cite{Ventura2006,Natarajan2010}.
\item Use of a quasi-polar integration with a $5\times5$ quadrature rule in triangular subdomains for elements containing the crack tip.
\item No correction for blending elements. Methods to address blending errors are proposed by \cite{Chessa2003,Gracie2008,Fries2008,Tarancon2009}.
\end{itemize}

\subsection{Evaluation of stress intensity factors}
\label{Sec:SIF}

The stress intensity factors (SIFs) in LEFM represent the amplitude of the singular stress fields and are key quantities of interest to simulate crack growth in LEFM. Several post--processing methods, following local or global (energy) approaches, are commonly used to extract SIFs \citep{Banks1991} or to calculate the energy release rate $G$. Energy or global methods are considered to be the most accurate and efficient methods \citep{Banks1991,Li1985}. Global methods based on the equivalent domain integral (EDI methods) are specially well-suited for FEM and XFEM analyses as they are easy to implement and can use information far from the singularity. In this paper, the interaction integral as described in \cite{Shih1988,Yau1980} has been used  to extract the SIFs. This technique provides $K_{\rm I}$ and $K_{\rm II}$ for problems under mixed mode loading conditions using auxiliary fields. Details on the implementation of the interaction integral can be found for example in \cite{Moes1999,Rodenas2007a,Giner2005}.

\section{Error estimation in the energy norm}

The approximate nature of the FEM and XFEM approximations implies a discretization error which can be quantified using the error in the energy norm  for the solution $\norm{\vm{e}}=\norm{\vm{u}-\vm{u}^h}$. To obtain an estimate of the discretization error $\norm{\vm{e}_{es}}$, in the context of elasticity problems solved using the FEM, the expression for the ZZ estimator is defined as \citep{Zienkiewicz1987} (in matrix form):

\begin{equation} \label{Eq:ZZ-estimator} 
\norm{\vm{e}}^{2} \approx \norm{\vm{e}_{es}}^{2}=\int _{\Omega}\left( \vm{\sigma}^*- \vm{\sigma}^h \right)^{T} \vm{D}^{-1} \left(\vm{\sigma}^*- \vm{\sigma}^h \right)d\Omega   
\end{equation}
\noindent or alternatively for the strains
\begin{equation}  
\norm{\vm{e}_{es}}^{2}=\int _{\Omega}\left( \vm{\varepsilon}^*- \vm{\varepsilon}^h \right)^{T} \vm{D} \left(\vm{\varepsilon}^*- \vm{\varepsilon}^h \right)d\Omega, \nonumber
\end{equation}

\noindent where the domain $\Omega$ refers to the full domain of the problem or a local subdomain (element), $\vm{\sigma}^h$ represents the stress field provided by the FEM, $\vm{\sigma}^*$ is the recovered stress field, which is a better approximation to the exact solution than $\vm{\sigma}^h$ and $\vm{D}$ is the elasticity matrix of the constitutive relation $\vm{\sigma} = \vm{D}\vm{\varepsilon}$.
   
The recovered stress field $\vm{\sigma}^*$ is usually interpolated in each element using the shape functions $\vm{N}$ of the underlying FE approximation and the values of the recovered stress field calculated at the nodes $\bar{\vm{\sigma}}^*$
 
\begin{equation} \label{Eq:rec-stress-interpolation}  
\vm{\sigma}^*(\vm{x}) = \sum _{i=1}^{n_e} N_{i}(\vm{x}) \bar{\vm{\sigma}}^*_i(\vm{x}_i),
\end{equation}

\noindent where $n_e$ is the number of nodes in the element under consideration and $\bar{\vm{\sigma}}^*_i(\vm{x}_i)$ are the stresses provided by the least squares technique at node $i$. The components of $\bar{\vm{\sigma}}^*_i$ are obtained using a polynomial expansion,  $\bar{\sigma}_{i,j} = \vm{p}\vm{a}$ (with $j=xx,yy,xy$), defined over a set of contiguous elements connected to node $i$ called \textit{patch}, where $\vm{p}$ is the polynomial basis and $\vm{a}$ are the unknown coefficients. 

The ZZ error estimator is asymptotically exact if the recovered solution used in the error estimation converges at a higher rate than the finite element solution \citep{Zienkiewicz1992b}. 

In this paper, we are interested in the role played by statical admissibility and enrichment of the recovered solution in estimating the error committed by XFEM. To do so, we study the performance of two recovery-based error estimators, which exhibit  different features, that have been recently developed for XFEM:
\begin{itemize}
\item The SPR-CX derived from the error estimator developed by  \cite{Rodenas2007a} and summarized in Section \ref{section:sprcx} (two other versions of the SPR-CX technique have been also considered);
\item The XMLS proposed by \cite{Bordas2007} and summarized in Section \ref{section:xmls}.
\end{itemize}

Both estimators provide a recovered stress field in order to evaluate the estimated error in the energy norm by means of the expression shown in (\ref{Eq:ZZ-estimator}).

\section{SPR-CX error estimator} \label{section:sprcx}

The SPR-CX error estimator is an enhancement of the error estimator first introduced by \cite{Rodenas2007a}, which incorporates the ideas proposed in \cite{Rodenas2007} to guarantee the exact satisfaction of the equilibrium locally on patches. In \cite{Rodenas2007a} a set of key ideas are proposed to modify the standard SPR by \cite{Zienkiewicz1992}, allowing its use for singular problems. 

The recovered stresses $\vm{\sigma}^*$ are directly evaluated at an integration point $\vm{x}$ through the use of a partition of unity procedure, properly weighting the stress interpolation polynomials obtained from the different patches formed at the vertex nodes of the element containing $\vm{x}$:

\begin{equation} \label{Eq:conjoint_polynomials} 
\vm{\sigma}^*(\vm{x}) = \sum _{i=1}^{n_v} N_{i}(\vm{x}) \vm{\sigma}^*_i(\vm{x}),
\end{equation}

\noindent where $N_{i}$ are the shape functions associated to the vertex nodes $n_v$.

One major modification is the introduction of a splitting procedure to perform the recovery. For singular problems the exact stress field $\vm{\sigma}$  is decomposed into two stress fields, a smooth field $\vm{\sigma}_{smo}$ and a singular field $\vm{\sigma}_{sing}$:

\begin{equation} \label{Eq:splitting} 
\vm{\sigma} =   \vm{\sigma}_{smo} + \vm{\sigma}_{sing}.
\end{equation}

Then, the recovered field $\vm{\sigma}^*$ required to compute the error estimate given in (\ref{Eq:ZZ-estimator}) can be expressed as the contribution of two recovered stress fields, one smooth $\vm{\sigma}^*_{smo}$ and one singular $\vm{\sigma}^*_{sing}$:

\begin{equation} \label{Eq:recovered_splitting} 
\vm{\sigma}^* =   \vm{\sigma}^*_{smo} + \vm{\sigma}^*_{sing}.
\end{equation}

For the recovery of the singular part, the expressions in (\ref{Eq:CrackTipField:Stress}) which describe the asymptotic fields near the crack tip are used. To evaluate $\vm{\sigma}^*_{sing}$ from  (\ref{Eq:CrackTipField:Stress}) we first obtain estimated values of the stress intensity factors $K_{\rm I}$ and $K_{\rm II}$ using the interaction integral as indicated in Section~\ref{Sec:SIF}. The recovered part $\vm{\sigma}^*_{sing}$ is an equilibrated field as it satisfies the internal equilibrium equations.

Once the field $\vm{\sigma}^*_{sing}$ has been evaluated, an FE approximation to the smooth part $\vm{\sigma}^h_{smo}$ can be obtained subtracting $\vm{\sigma}^*_{sing}$ from the raw FE field:

\begin{equation} \label{Eq:tens_h_smooth} 
\vm{\sigma}^h_{smo} =   \vm{\sigma}^h - \vm{\sigma}^*_{sing}.
\end{equation}

Then, the field $\vm{\sigma}^*_{smo}$ is evaluated applying an SPR-C recovery procedure over the field $\vm{\sigma}^h_{smo}$.

For patches intersected by the crack, the recovery technique uses different stress interpolation polynomials on each side of the crack. This way it can represent the discontinuity of the recovered stress field along the crack faces, which is not the case for SPR that smoothes out the discontinuity (see, e.g. \cite{Bordas2007}).

In order to obtain an equilibrated recovered stress field $\vm{\sigma}^*_{smo}$, the SPR-CX enforces the fulfilment of the equilibrium equations locally on each patch. The constraint equations are introduced via Lagrange multipliers into the linear system used to solve for the coefficients of the polynomial expansion of the recovered stresses on each patch. These include the satisfaction of the:

\begin{itemize}
	\item Internal equilibrium equations.
	\item Boundary equilibrium equation: A point collocation approach is used to impose the satisfaction of a second order approximation to the tractions along the Neumann boundary.  
	\item Compatibility equation: This additional constraint is also imposed to further increase the accuracy of the recovered stress field.  
\end{itemize}

To evaluate the recovered field, quadratic polynomials have been used in the patches along the boundary and crack faces, and linear polynomials for the remaining patches. As more information about the solution is available along the boundary, polynomials one degree higher are useful to improve the quality of the recovered stress field.

The enforcement of equilibrium equations provides an equilibrated recovered stress field locally on patches. However, the process used to obtain a continuous field $\vm{\sigma}^*$ shown in (\ref{Eq:conjoint_polynomials}) introduces a small lack of equilibrium as explained in \cite{Rodenas2010}. The reader is referred to \cite{ Rodenas2010, Diez2007} for details.

\section{XMLS error estimator} \label{section:xmls}

In the XMLS the solution is recovered through the use of the \emph{moving least squares} (MLS) technique, developed by mathematicians to build and fit surfaces. The XMLS technique extends the work of \cite{Tabbara1994} for FEM to enriched approximations. The general idea of the XMLS is to use the displacement solution provided by XFEM to obtain a recovered strain field \citep{Bordas2007, Bordas2008}. The smoothed strains are recovered from the derivative of the MLS-smoothed XFEM displacement field:
\red{
\begin{align}
	\vm{u} ^* (\vm{x}) &= \sum_{i \in \mathcal{N}_x} \psi _{i} (\vm{x}) \vm{u} _{i}^{h} \label{Eq:XMLSdisp} \\
	\vm{\varepsilon} ^* (\vm{x} )&= \sum_{i \in \mathcal{N}_x} \nabla_s  (\psi _{i} (\vm{x}) \vm{u} _{i}^{h} ) \label{Eq:XMLSstrain}, 
\end{align}
}

\noindent \red{where the $\vm{u} _{i}^{h}$ are the raw nodal XFEM displacements and $\psi _{i}(\vm{x})$ are the MLS shape function values associated with a node $i$ at a point $\vm{x}$. $ \mathcal{N}_x$ is the set of $n_x$ nodes in the domain of influence of point $\vm{x}$, $\nabla_s$ is the symmetric gradient operator, $\vm{u} ^*$ and $\vm{\varepsilon} ^*$ are the recovered displacement and strain fields respectively. At each point $\vm{x}_i$ the MLS shape functions $\psi _{i}$ are evaluated using weighting functions $\omega_i$ and an enriched basis $\vm{p}(\vm{x}_i)$. The total $n_x$ non-zero MLS shape functions at point $\vm{x}$ are evaluated as:
\begin{equation}
 (\psi _{i} (\vm{x}))_{1\leq i \leq n_x} = (\vm{A}^{-1}(\vm{x})\vm{p} (\vm{x}))^T  \vm{p}(\vm{x}_i) \omega_i (\vm{x})
\end{equation}
\noindent where $\vm{A}(\vm{x})= \sum _{i=1}^{n_x} \omega_i (\vm{x}) \vm{p}(\vm{x}_i)\vm{p}(\vm{x}_i)^T$ is a matrix to be inverted at every point $\vm{x}$ (see \cite{Bordas2007, Bordas2008} for further details).  
}

\begin{figure}[!ht]
	\centering
		\includegraphics[scale=1]{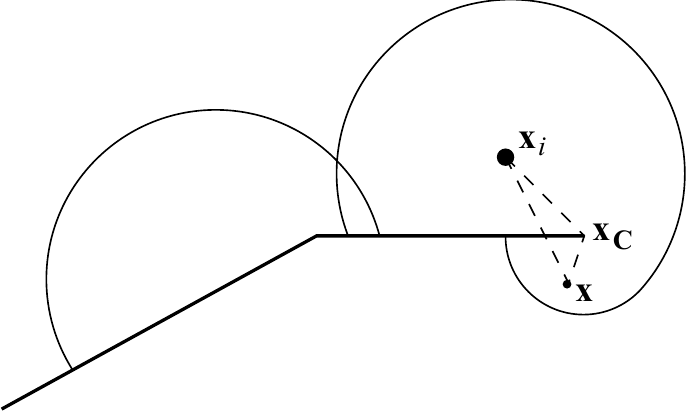}
	\caption{Diffraction criteria to introduce the discontinuity in the XMLS approximation.}
	\label{fig:XMLS_diffraction}
\end{figure}

For each supporting point $\vm{x}_i$, the weighting function $\omega_i$ is defined such that:
\begin{equation}
	\omega_i(s) = f_4 (s) = 
	\begin{cases}
	1- 6s^2+8s^3-3s^4 &\text{if }\left|s\right|\leq 1\\
	0	                &\text{if }\left|s\right|> 1
	\end {cases}
\end{equation}

\noindent where $s$ is the normalized distance between the supporting point $\vm{x}_i$ and a point $\vm {x}$ in the computational domain. In order to describe the discontinuity, the distance $s$ in the weight function defined for each supporting point is modified using the \emph{diffraction criterion} \citep{Belytschko1996a}. The basic idea of this criterion is depicted in Figure~\ref{fig:XMLS_diffraction}. The weight function is continuous except across the crack faces since the points at the other side of the crack are not considered as part of the support. \red{Near the crack tip the weight of a node $i$ over a point of coordinates $\vm{x}$ diminishes as the crack hides the point. When the point $\vm{x}$ is hidden by the crack the following expression is used}:
\begin{equation}\label{Eq:XMLSdifraction} 
	s=\frac{\left\|\vm{x}-\vm{x}_C\right\| + \left\| \vm{x}_C-\vm{x}_i \right\|}{d_i}
\end{equation}

\noindent where $d_i$ is the radius of the support.

The MLS shape functions can reproduce any function in their basis. The basis $\vm{p}$ used is a linear basis enriched with the functions that describe the first order asymptotic expansion at the crack tip as indicated in \eqref{Eq:branch_functions}:

\begin{equation}\label{Eq:XMLSbasis} 
	\vm{p}=\left[1,x,y,\left[ F_1(r,\phi),F_2(r,\phi),F_3(r,\phi),F_4(r,\phi)\right]\right]
\end{equation}

Note that although the enriched basis can reproduce the singular behaviour of the solution around the crack tip, the resulting recovered field not necessarily satisfies the equilibrium equations.

\section{Numerical results}

In this section, numerical experiments are performed to verify the behaviour of both XFEM recovery based error estimators considered in this paper.

\cite{Babuska1994,Babuska1994b,Babuska1997} proposed a robustness patch test for quality assessment of error estimators. However, the use of this test is not within the scope of this paper and furthermore, to the authors' knowledge, it has not been used in the context of XFEM. Therefore, the more traditional approach of using benchmark problems is considered here to analyse the response of the different error estimators.

The accuracy of the error estimators is evaluated both locally and globally. This evaluation has been based on the effectivity of the error in the energy norm, which is quantified using the \emph{effectivity index} $\efec$ defined as:

\begin{equation} \label{Eq:Effect_Index} 
\efec =\frac{\normees }{\norme}.
\end{equation}

When the enhanced or recovered solution is close to the analytical solution the effectivity approaches the theoretical value of 1, which indicates that it is a good error estimator, i.e. the approximate error is close to the exact error.

To assess the quality of the estimator at a local level, the local effectivity $\efecl$, inspired on the \emph{robustness index} found in \cite{Babuska1994}, is used. For each element $e$, $\efecl$ represents the variation of the effectivity index in this element, $\efece$, with respect to the theoretical value (the error estimator can be considered to be of good quality if it yields $\efecl$ values close to zero). $\efecl$ is defined 
according to the following expression, where superscripts $^{e}$ indicate the element $e$: 

\begin{equation} \label{Eq:Local_Effect} 
\begin{array}{c} {
\displaystyle \efecl=\efece -1\qquad {\rm if}\qquad \efece \ge  1} \\ 
{\displaystyle \efecl=1-\frac{1}{\efece } \qquad {\rm if}\qquad \efece < 1} 
\end{array}
\qquad \qquad {\rm with}\qquad \efece =\frac{\left\| \errore_{es}^{e} \right\| }{\left\| \errore^{e} \right\| }.
\end{equation}
	
To evaluate the overall quality of the error estimator we use the global effectivity index $\efec$, the mean value $\meanD$ and the standard deviation $\stdD$ of the local effectivity. A good quality error estimator yields  values of $\efec$ close to one and values of $\meanD$ and $\stdD$ close to zero.

\red{The techniques can be used in practical applications. However, in order to properly compare their performance we have used an academic problem with exact solution}. In the analysis we solve the Westergaard problem \citep{gdoutos1993} as it is one of the few problems in LEFM under mixed mode with an analytical solution. In \cite{Giner2005,Rodenas2007a} we can find explicit expressions for the stress fields in terms of the spatial coordinates. In the next subsection we show a description of the Westergaard problem and XFEM model, taken from \cite{Rodenas2007a} and reproduced here for completeness.

\subsection{Westergaard problem and XFEM model}

\begin{figure}[ht]
	\centering
		 \includegraphics{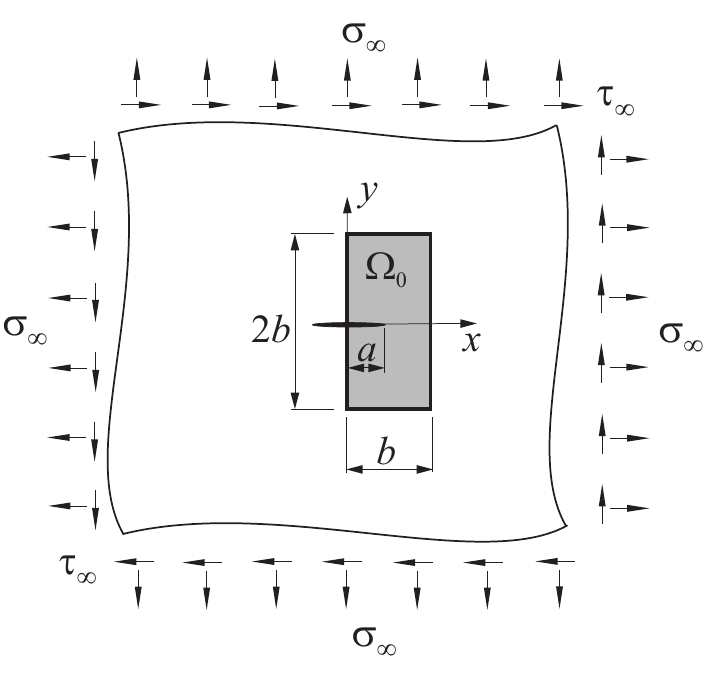}
	\caption{Westergaard problem. Infinite plate with a crack of length $2a$ under uniform tractions $\sigma_{\infty}$ (biaxial) and $\tau_{\infty}$. Finite portion of the domain $\Omega_0$, modelled with FE.}
	\label{fig:westergaard}
\end{figure}

The Westergaard problem corresponds to an infinite plate loaded at infinity with biaxial tractions $\sigma_{x \infty}=\sigma_{y \infty}=\sigma_{\infty}$ and shear traction $\tau_{\infty}$, presenting a crack of length $2a$ as shown in Figure~\ref{fig:westergaard}. Combining the externally applied loads we can obtain different loading conditions: pure mode I, II or mixed mode.  

In the numerical model only a finite portion of the domain ($a=1$ and $b=4$ in Figure~\ref{fig:westergaard}) is considered. The projection of the stress distribution corresponding to the analytical Westergaard solution for modes I and II, given by the expressions below, is applied to its boundary:

\begin{equation} \label{Eq:westergaard_stress_I} 
\begin{array}{r@{\hspace{1ex}}c@{\hspace{1ex}}l} {\sigma _{x}^{I} } (x,y) & {=} & {\displaystyle \frac{\sigma _{\infty } }{\sqrt{\left|t\right|} } \bigg[\left(x\cos \frac{\phi }{2} -y\sin \frac{\phi }{2} \right)+y\frac{a^{2} }{\left|t\right|^{2} } \left(m\sin \frac{\phi }{2} -n\cos \frac{\phi }{2} \right)\bigg]} \\ \noalign{\medskip}{\sigma _{y}^{I} }(x,y) & {=} & {\displaystyle \frac{\sigma _{\infty } }{\sqrt{\left|t\right|} } \bigg[\left(x\cos \frac{\phi }{2} -y\sin \frac{\phi }{2} \right)-y\frac{a^{2} }{\left|t\right|^{2} } \left(m\sin \frac{\phi }{2} -n\cos \frac{\phi }{2} \right)\bigg]} \\ \noalign{\medskip}{ \tau _{xy}^{I} }(x,y) & {=} & {\displaystyle y\frac{a^{2} \sigma _{\infty } }{\left|t\right|^{2} \sqrt{\left|t\right|} } \left(m\cos \frac{\phi }{2} +n\sin \frac{\phi }{2} \right)} \end{array}
\end{equation}

\begin{equation} \label{Eq:westergaard_stress_II} 
\begin{array}{r@{\hspace{1ex}}c@{\hspace{1ex}}l} {\sigma _{x}^{II}}(x,y)  & {=} & {\displaystyle \frac{\tau _{\infty } }{\sqrt{\left|t\right|} } \bigg[2\left(y\cos \frac{\phi }{2} +x\sin \frac{\phi }{2} \right)-y\frac{a^{2} }{\left|t\right|^{2} } \left(m\cos \frac{\phi }{2} +n\sin \frac{\phi }{2} \right)\bigg]} \\ \noalign{\medskip}{\sigma _{y}^{II}}(x,y)  & {=} & {\displaystyle y\frac{a^{2} \tau _{\infty } }{\left|t\right|^{2} \sqrt{\left|t\right|} } \left(m\cos \frac{\phi }{2} +n\sin \frac{\phi }{2} \right)} \\ \noalign{\medskip}{\tau _{xy}^{II}}(x,y)  & {=} & {\displaystyle \frac{\tau _{\infty } }{\sqrt{\left|t\right|} } \bigg[\left(x\cos \frac{\phi }{2} -y\sin \frac{\phi }{2} \right)+y\frac{a^{2} }{\left|t\right|^{2} } \left(m\sin \frac{\phi }{2} -n\cos \frac{\phi }{2} \right)\bigg]} \end{array}
\end{equation}

\noindent where the stress fields are expressed as a function of $x$ and $y$, with origin at the centre of the crack. The parameters $t$, $m$, $n$ and $\phi$ are defined as

\begin{equation}
\begin{split}t& =(x+iy)^{2} -a^{2} =(x^{2} -y^{2} -a^{2} )+i(2xy)=m+in \\  m & =\textrm{Re}(t) =\textrm{Re}(z^{2} -a^{2} )=x^{2} -y^{2} -a^{2} \\ n & =\textrm{Im}(t)=(z^{2} -a^{2} )=2xy \\ \phi & =\textrm{Arg} (\bar{t})=\textrm{Arg} (m-in) \qquad\textrm{with }\phi \in \left[-\pi ,\pi \right], \; i^2=-1 \end{split} 
\end{equation}

For the problem analysed, the exact value of the SIF is given as 

\begin{equation} \label{Eq:SIFWestergaard} 
K_{{\rm I},ex} =\sigma \sqrt{\pi a} \qquad \qquad K_{{\rm II},ex} =\tau \sqrt{\pi a}  
\end{equation}

\begin{figure}[!ht]
	\centering
		 \includegraphics{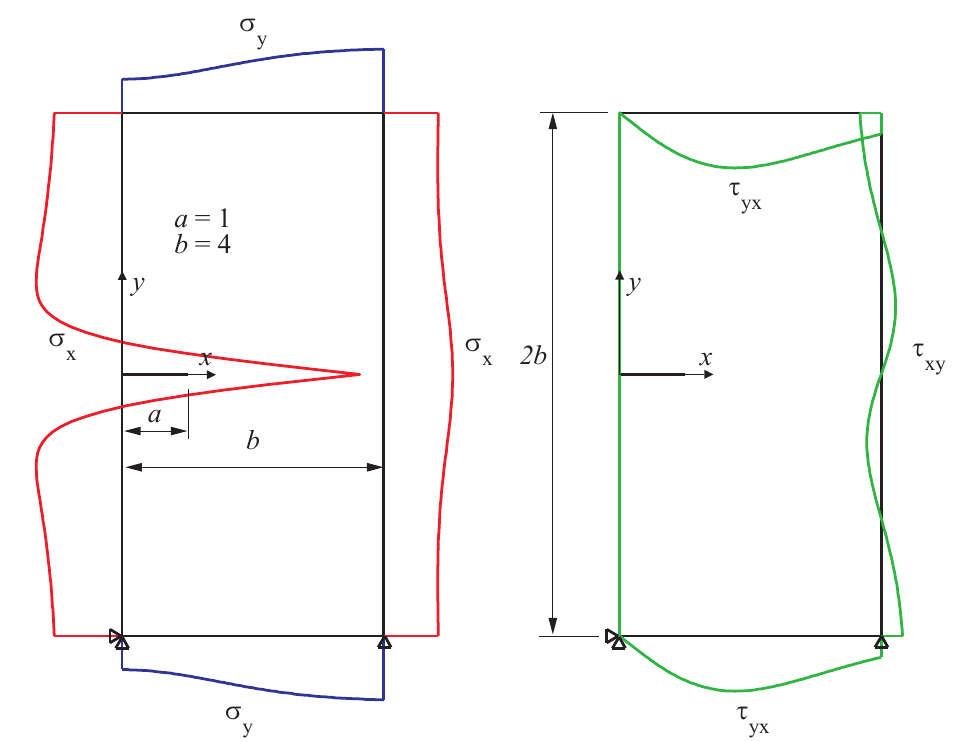}
	\caption{Model for an infinite plate with a crack subjected to biaxial tractions $\sigma _{\infty },\;\tau _{\infty }$ in the infinite.}
	\label{fig:WestergaardPlate}
\end{figure}

Three different loading configurations corresponding to the \emph{pure mode I} ($\sigma _{\infty }=100,\;\tau _{\infty }=0$), \emph{pure mode II} ($\sigma _{\infty }=0,\;\tau _{\infty }=100$), and \emph{mixed mode} ($\sigma _{\infty }=30,\;\tau_{\infty }=90$) cases of the Westergaard problem are considered. The geometric models and boundary conditions  are shown in Figure~\ref{fig:WestergaardPlate}.

Bilinear elements are used in the models, with a \emph{singular}+\emph{smooth} decomposition area of radius  $\rho = 0.5$ equal to the radius $r_e$ used for the fixed enrichment area \red{(note that the splitting radius $\rho$ should be greater or equal to the enrichment radius $r_e$, \cite{Rodenas2007a})}. The radius for the Plateau function used to extract the SIF is $r_q = 0.9$. Young's modulus is $E = 10^7$ and Poisson's ratio $ \upsilon= 0.333$. For the numerical integration of standard elements we use a $2\times2$ Gaussian quadrature rule. For split elements we use 7 Gauss points in each triangular subdomain, and a $5\times 5$ quasipolar integration  \citep{Bechet2005} in the subdomains of the element containing the crack tip. 

\begin{figure}[!ht]
	\centering
		\includegraphics[width=1.00\textwidth]{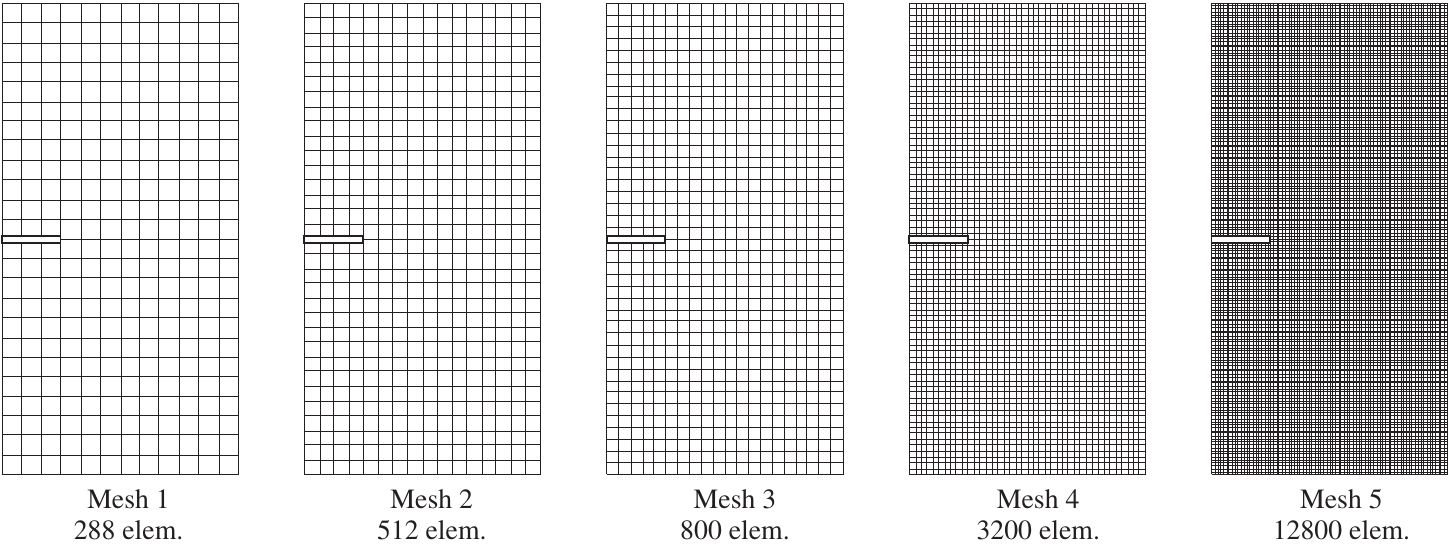}
	\caption{Sequence of structured meshes.}
	\label{fig:struct_mesh}
\end{figure}

\begin{figure}[!ht]
	\centering
		\includegraphics[width=1.00\textwidth]{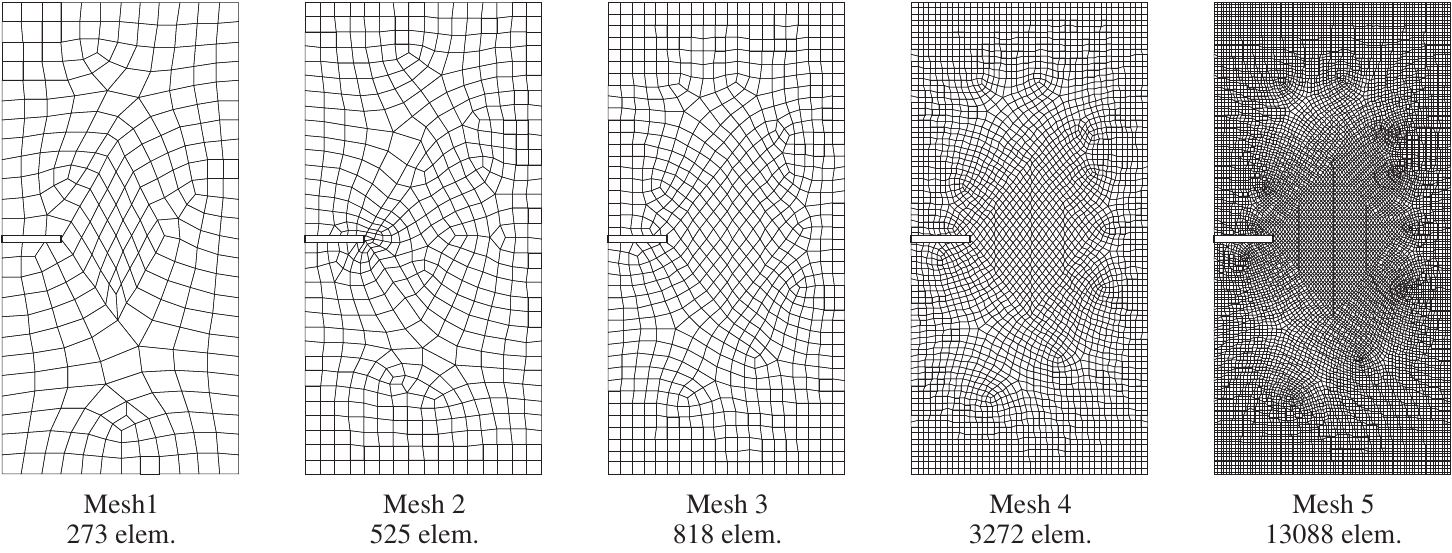}
	\caption{Sequence of unstructured meshes.}
	\label{fig:n-struct_mesh}
\end{figure}

Regarding global error estimation, the evolution of global parameters in sequences of uniformly refined structured (Figure~\ref{fig:struct_mesh}) and unstructured (Figure~\ref{fig:n-struct_mesh}) meshes is studied. In the first case, the mesh sequence is defined so that the crack tip always coincided with a node. For a more general scenario, this condition has not been applied for the unstructured meshes.

\subsection{Mode I and structured meshes}

\begin{figure}[!ht]
	\centering
		\includegraphics[scale=0.85]{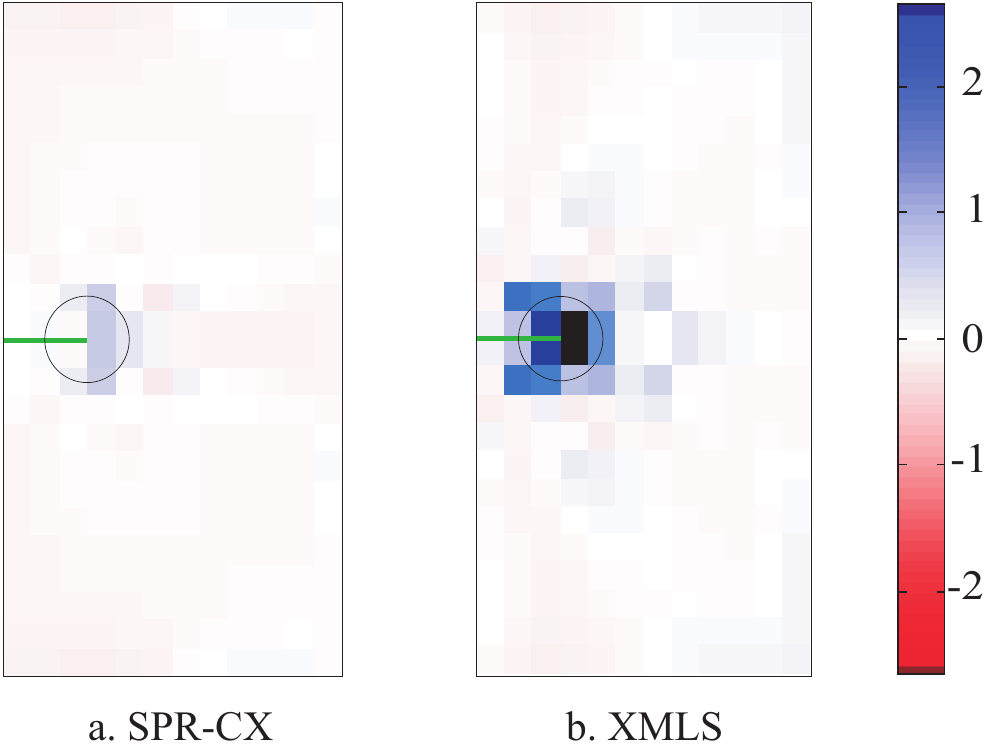}
	\caption{Mode I, structured mesh 1.  Local effectivity index $\efecl$ (ideal value $\efecl=0$). The circles denote the enriched zones around the crack tips. The superiority of the statically admissible recovery (SPR-CX) compared to the standard XMLS is clear. }
	\label{fig:XMLSvsSPRCX_M1_D}
\end{figure}

The first set of results presented in this study are for the Westergaard problem under mode I load conditions. Figure~\ref{fig:XMLSvsSPRCX_M1_D}  shows the values for the local effectivity index $\efecl$ for the first mesh in the sequence analysed. In the figure, the size of the enrichment area is denoted by a circle. It can be observed that far from the enrichment area both techniques yield similar results, however, the XMLS technique exhibits a higher overestimation of the error close to the singularity.

\begin{figure}[!ht]
	\centering
		\includegraphics[scale=0.85]{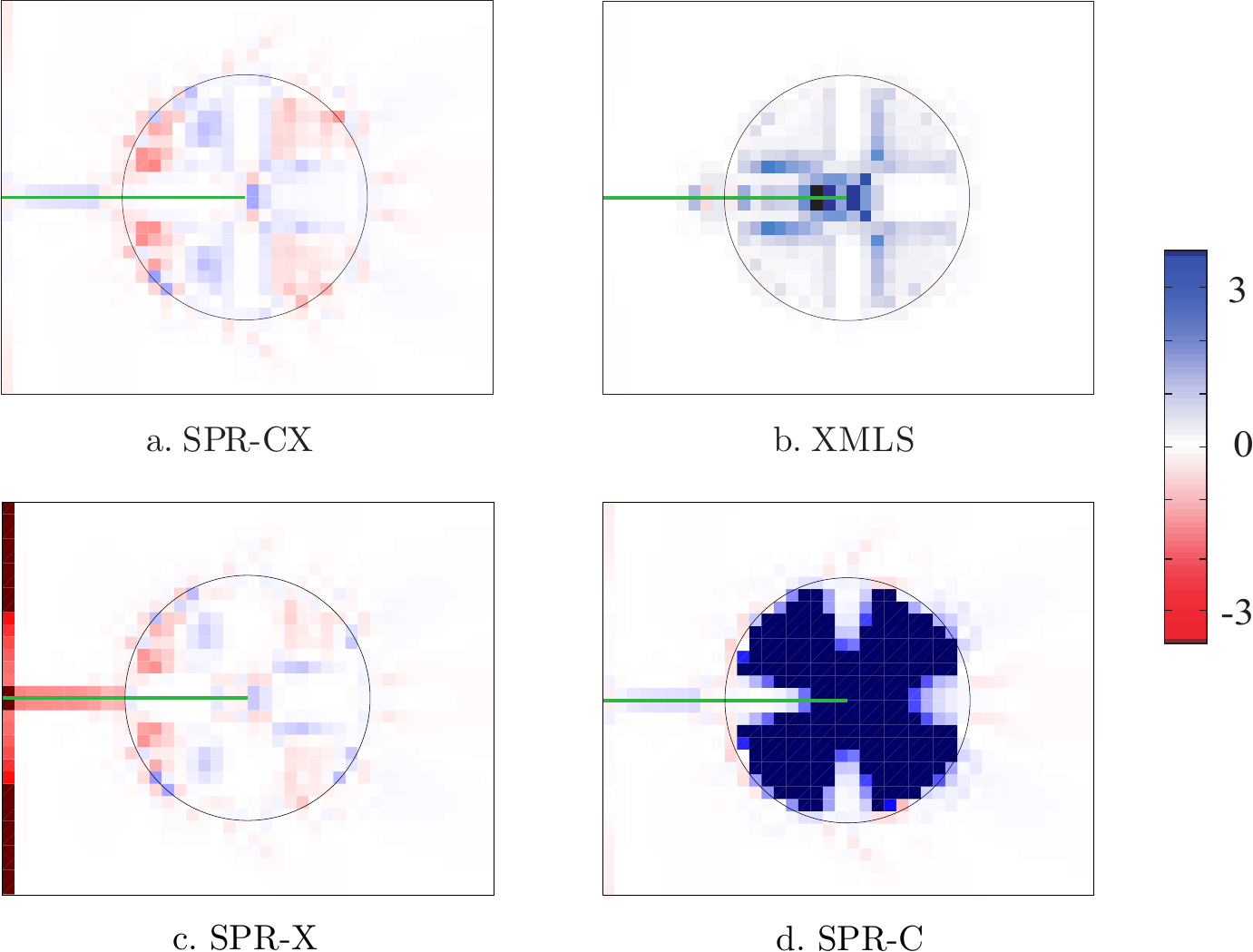}
	\caption{Mode I, structured mesh 4. Local effectivity index $\efecl$ (ideal value $\efecl=0$). The circles denote the enriched zones around the crack tips. The results show the need for the inclusion of the near-tip fields in the recovery process (SPR-X is far superior to SPR-C). It is also clear that enforcing statical admissibility (SPR-CX) greatly improves the accuracy along the crack faces and the Neumann boundaries.  Also notice that XMLS leads to slightly improved effectivities compared to SPR-CX around the crack faces, between the enriched zone and the left boundary. SPR-C is clearly not able to reproduce the singular fields, which is shown by large values of $D$ inside the enriched region. }
	\label{fig:XMLSvsSPRCX_M1_Dzoom}
\end{figure}

The same behaviour is also observed for more refined meshes in the sequence. Figure~\ref{fig:XMLSvsSPRCX_M1_Dzoom}  shows a zoom in the enriched area of a finer mesh where, as before, a higher overestimation of the error can be observed for the results obtained with the XMLS error estimator. In this example and in order to study the evolution of the accuracy of the recovered stress field when considering different features in the recovery process, we consider two additional versions:
\begin{itemize}
\item The first  considers a recovery procedure which enforces both internal and boundary equilibrium, but does not include the \textit{singular+smooth} splitting technique (SPR-C). This approach is very similar in form to conventional SPR-based recovery techniques widely used in FEM. It is well-known that this type of recovery process will produce unreliable results when the stress field contains a singularity because the polynomial representation of the recovered stresses is not able to describe the singular field (e.g. \cite{Bordas2007}). 
\item The second version of the error estimator performs the splitting but does not equilibrate the recovered field (SPR-X). As previously commented, $\vm{\sigma}^*_{sing}$ is an equilibrated field but $\vm{\sigma}^*_{smo}$ is not equilibrated. The aim of this second version is to assess the influence of enforcing the equilibrium constraints. 
\end{itemize}
Table 1 summarizes the main features of the different recovery procedures we considered.

\begin{table} [!ht] 
    \begin{center}
	\begin{tabular}{|c|c|c|c|}
	\hline  & Singular &  Equilibrated sin-    & Locally equili-   \\ 
	        & functions &  gular functions     &  brated field \\
	\hline SPR-X 	& yes 	& yes 	& no \\ 
	\hline SPR-C 	& no 	&  not applicable 	& yes\\ 
	\hline SPR-CX 	& yes 	& yes 	& yes \\ 
	\hline XMLS	    & yes 	& no 	& no\\ 
	\hline 
	\end{tabular} 
	\caption{Comparison of features for the different recovery procedures}
	\end{center}
	\label{Tab:Features}
\end{table}

The performance of SPR-X in Figure~\ref{fig:XMLSvsSPRCX_M1_Dzoom} is poor, especially along the Neumann boundaries, where the equilibrium equations are not enforced, as they are in the SPR-CX. Particularly interesting are the results for the SPR-C technique, where considerable overestimation of the exact error is observed \emph{in the whole enriched region}. This is due to the inability of polynomial functions to reproduce the near-tip fields, even relatively far from the crack tip.  

\begin{figure}[!ht]
	\centering
	 \includegraphics{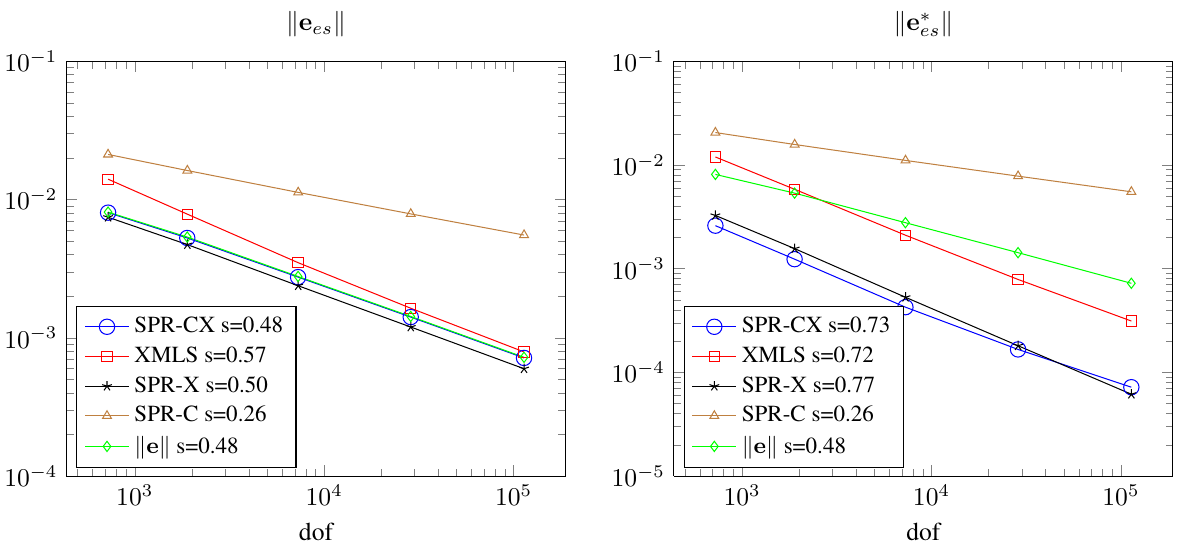}
	\caption{Mode I, structured meshes: Convergence of the estimated error in the energy norm $\Vert \mathbf{e} _{es}\Vert$ (left). All methods which include the near-tip fields do converge with the optimal convergence rate of 1/2. The best method is SPR-CX. XMLS performs worse than SPR-X for coarse meshes, but it becomes equivalent to SPR-CX for fine meshes whilst the SPR-X error increases slightly.  Convergence of the error for the recovered field $\Vert \mathbf{e} _{es}^*\Vert$ (right). For the recovered field, SPR-CX is the best method closely followed by SPR-X. XMLS errors are half an order of magnitude larger than SPR-CX/SPR-X, but superior to the non-enriched recovery method SPR-C. For all methods, except SPR-C, the error on the recovered solution converges faster than the error on the raw solution, which shows that the estimators are asymptotically exact.}
	\label{fig:XMLSvsSPRCX_M1_enorm}
\end{figure}

The above  compared the relative performance of the methods locally. It is also useful to have a measure of the global accuracy of the different error estimators, Figure~\ref{fig:XMLSvsSPRCX_M1_enorm}(left) shows the convergence of the estimated error in the energy norm $\normees$ (see equation (\ref{Eq:ZZ-estimator})) evaluated with the proposed techniques, the convergence of the exact error $\norme$\footnote{The exact error measures the error of the raw XFEM compared to the exact solution. The theoretical convergence rate is $\mathcal{O}(h^p)$, $h$ being the element size and $p$ the polynomial degree (1 for linear elements,...), or $\mathcal{O}(-\text{dof}^{p/2})$ if we consider the number of  degrees of freedom.} is shown for comparison. It can be observed that SPR-CX leads to the best approximation to the exact error in the energy norm. The best approach is that which captures as closely as possible the exact error. Non-singular formulations such as SPR-C greatly overestimate the error, especially close to the singular point. Only SPR-C suffers from the characteristic suboptimal convergence rate (factor 1/2, associated with the strength of the singularity)

In order to evaluate the quality of the recovered field $\vm{\sigma}^*$ obtained with each of the recovery techniques, the convergence in energy norm of the approximate error on the recovered field $\normestar = \norm{\vm{u}-\vm{u}^*}$ is compared for the error estimators in Figure~\ref{fig:XMLSvsSPRCX_M1_enorm}(right). It can be seen that the XMLS, the SPR-CX and the SPR-X provide convergence rates higher than the convergence rate for the exact error $\norme$, which is an indicator of the asymptotic exactness of the error estimators based on these recovery techniques \citep{Zienkiewicz1992b}.

\begin{figure}[ht]
	\centering
		 \includegraphics{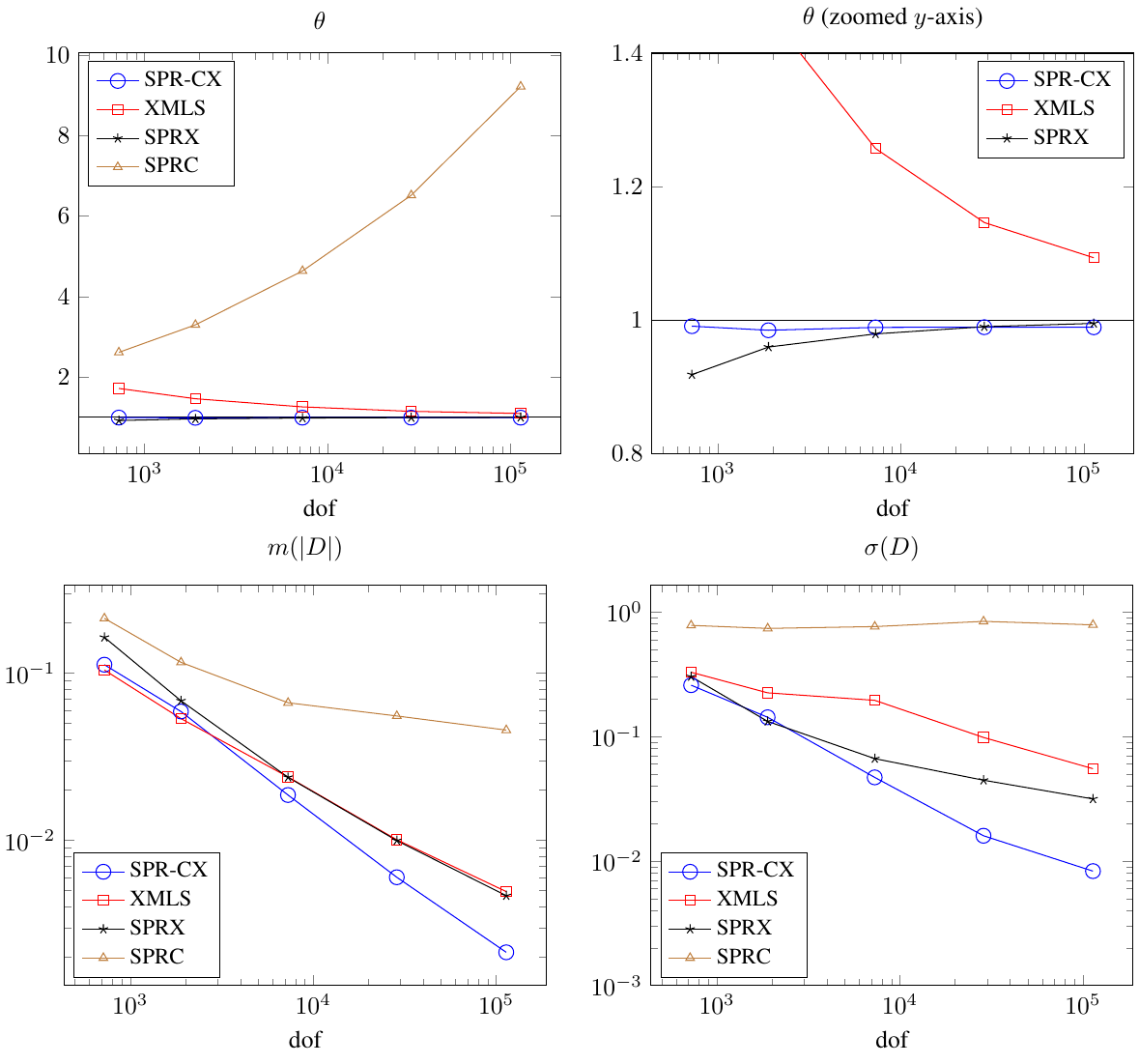}
	\caption{Global indicators $\efec$, $\meanD$ and $\stdD$ for mode I and structured meshes.}
	\label{fig:XMLSvsSPRCX_M1_Effec}
\end{figure}

Figure~\ref{fig:XMLSvsSPRCX_M1_Effec} shows the evolution of the parameters $\efec$, $\meanD$ y $\stdD$ with respect to the number of degrees of freedom for the error estimators.  It can be verified that although the SPR-C technique includes the satisfaction of the equilibrium equations, the effectivity of the error estimator does not converge to the theoretical value ($\theta=1$). This is due to the absence of the near-tip fields in the recovered solution. The influence of introducing singular functions in the recovery process can be observed in the results provided by SPR-X. In this case, the convergence is obtained both locally and globally. This shows the importance of introducing a description of the singular field in the recovery process, i.e. of using \emph{extended recovery techniques} in XFEM.

Regarding the enforcement of the equilibrium equations, we can see that an equilibrated formulation (SPR-CX) leads to better effectivities compared to other non-equilibrated configurations (SPR-X, XMLS), which is an indication of the advantages associated with equilibrated recoveries in this context.  

Still in Figure~\ref{fig:XMLSvsSPRCX_M1_Effec}, the results for the SPR-CX and XMLS show that the values of the global effectivity index are close to (and less than) unity for the SPR-CX estimator, whilst the XMLS is a lot less effective and shows effectivities larger than 1. 

The next step in our analysis is the verification of the convergence of the mean value and standard deviation of the local effectivity index $\efecl$. From a practical perspective, one would want the local effectivity to be equally good, on average, everywhere inside the domain; one would also want this property to improve with mesh refinement. In other words, the average local effectivity $m$ and its standard deviation $\sigma$ should decrease with mesh refinement: $m$ and $\sigma$ measure the average behaviour of the method and how far the results deviate from the mean, i.e. how spatially consistent they are. The idea is to spot areas where there may be compensation between overestimated and underestimated areas that could produce an apparently 'accurate' global estimation. It can be confirmed that, although the curves for $m(|D|)$ and $\sigma(D)$ for both estimators tend to zero with mesh refinement, the SPR-CX is clearly superior to the XMLS. These results can be explained by the fact that the SPR-CX enforces the fulfilment of the equilibrium equations in patches, and evaluates the singular part of the recovered field using the equilibrated expression that represents the first term of the asymptotic expansion, whereas the XMLS enrichment uses only a set of singular functions that would be able to reproduce this term but are not necessarily equilibrated. 

Although the results for the SPR-CX proved to be more accurate, the XMLS-type techniques will prove useful to obtain error bounds. There is an increasing interest in evaluating upper and lower bounds of the error for XFEM approximations. Some work has been already done to obtain upper bounds using recovery techniques as indicated in \cite{Rodenas2010}, where the authors showed that an upper bound can be obtained if the recovered field is continuous and equilibrated. However, they demonstrated that due to the use of the conjoint polynomials process to enforce the continuity of $\vm{\sigma}^*$, some residuals in the equilibrium equations appear, and the recovered field is only nearly equilibrated. Then, correction terms have to be evaluated to obtain the upper bound. An equilibrated version of the XMLS could provide a recovered stress field which would be continuous and equilibrated, thus facilitating the evaluation of the upper bound. Initial results for this class of techniques can be found in \cite{Rodenas2009, Rodenas2010a}. 

In this first example we have shown the effect of using the SPR-C and SPR-X recovery techniques in the error estimation. Considering that these two techniques are special cases of the SPR-CX with inferior results, for further examples we will focus only on the SPR-CX and XMLS techniques.

\subsection{Mode II and structured meshes}

\begin{figure}[!ht]
	\centering
		\includegraphics[width=1.00\textwidth]{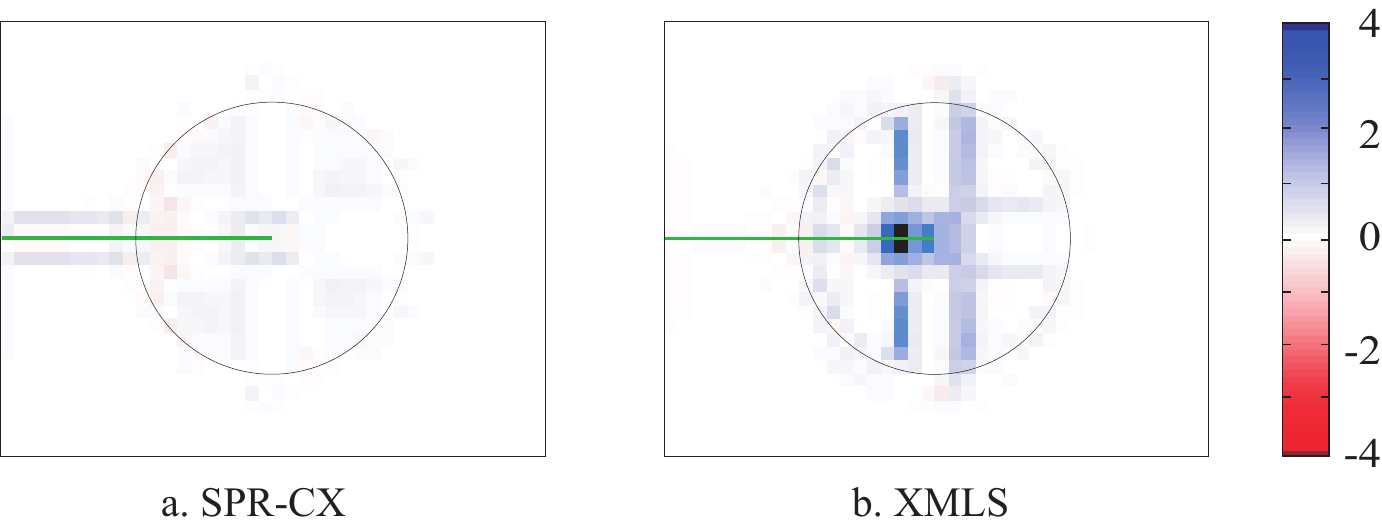}
	\caption{Mode II, structured mesh 4. Local effectivity index $\efecl$ (ideal value $\efecl=0$). SPR-CX leads to much better local effectivities than XMLS. It is also remarkable that the effectivities are clearly worse in the enriched region (circle) for both the XMLS and the SPR-CX, and that this effect is more pronounced in the former method. Note the slightly worse results obtained by SPR-CX around the crack faces between the boundary of the enriched region and the left boundary, as in mode I.  }
	\label{fig:XMLSvsSPRCX_M2_Dzoom}
\end{figure}

Figure~\ref{fig:XMLSvsSPRCX_M2_Dzoom} presents the results considering mode II loading conditions for the local effectivity index $\efecl$ on the fourth mesh of the sequence. Similarly to the results for mode I, the XMLS estimator presents a higher overestimation of the error near the enrichment area. This same behaviour is observed for the whole set of structured meshes analysed under pure mode II.

\begin{figure}[!ht]
\centering
	 \includegraphics{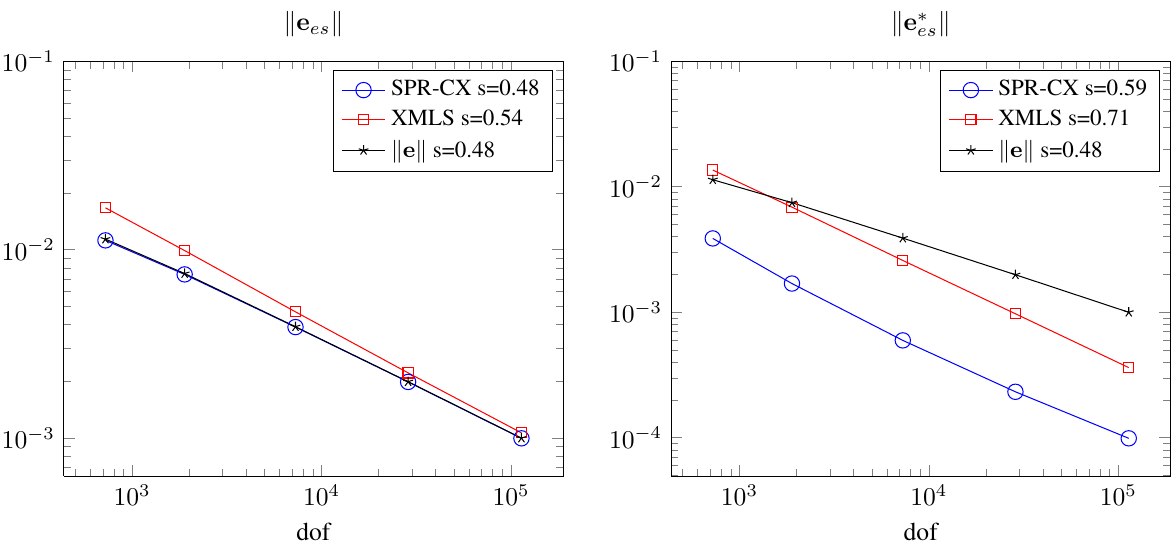}
	\caption{Mode II, structured meshes:  Convergence of the estimated error in the energy norm $\Vert \mathbf{e} _{es}\Vert$ (left).  The convergence rates are very similar to those obtained in the mode I case, and, SPR-CX is still the most accurate method, i.e. that for which the estimated error is closest to the exact error. Convergence of the error in the recovered field $\Vert \mathbf{e} _{es}^*\Vert$ (right). It can be noticed that the SPR-CX error  still converges faster than the exact error, but with a lower convergence rate (0.59 versus 0.73) than in the mode I case. There is no such difference for the XMLS results, which converge at practically the same rate (0.71 versus 0.72). The error level difference between SPR-CX and XMLS is about half an order of magnitude, as in the mode I case. }
	\label{fig:XMLSvsSPRCX_M2_enorm}
\end{figure}

\begin{figure}[!ht]
	\centering
	 \includegraphics{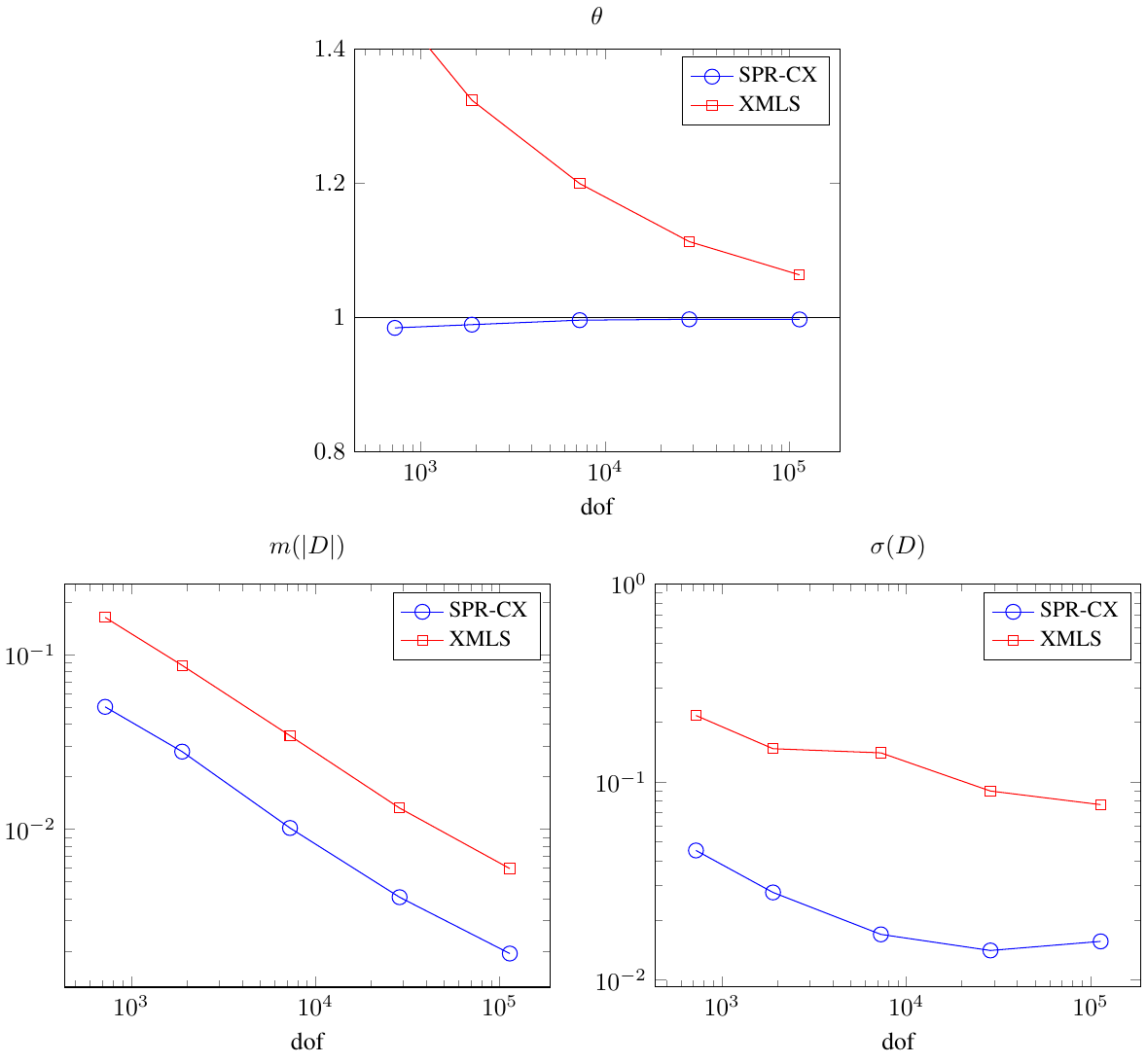}
	\caption{Global indicators $\efec$, $\meanD$ and $\stdD$ for mode II and structured meshes. The results are qualitatively and quantitatively similar to those obtained in mode I.}
	\label{fig:XMLSvsSPRCX_M2_Effec}
\end{figure}

As for the mode I case, the evolution of global accuracy parameters in mode II exhibits the same behaviour seen for mode I loading conditions. Figure~\ref{fig:XMLSvsSPRCX_M2_enorm} shows the results for the convergence of the error in the energy norm. Figure~\ref{fig:XMLSvsSPRCX_M2_Effec} shows the evolution of global indicators $\efec$, $\meanD$ and $\stdD$. Again, SPR-CX provides the best results for the Westergaard problem under pure mode II, using structured meshes. 

\subsection{Mixed mode and unstructured meshes}

\begin{figure}[!ht]
	\centering
		\includegraphics[width=1.00\textwidth]{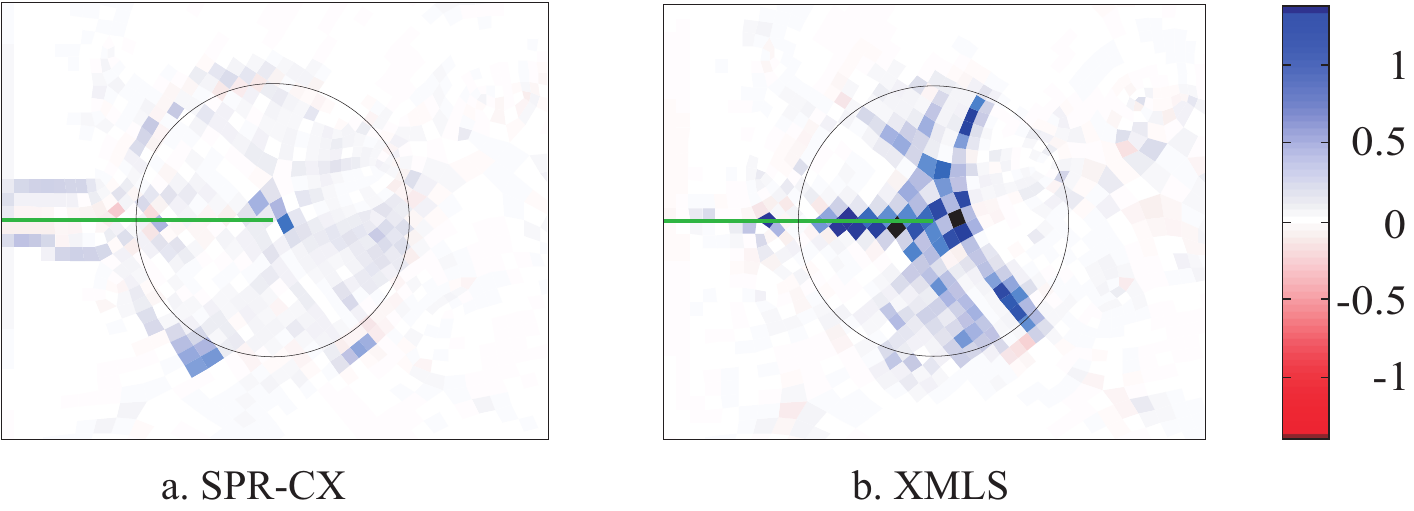}
	\caption{Mixed mode, unstructured mesh 4. Local effectivity index $\efecl$ (ideal value $\efecl=0$). Note the improved results compared to the structured case with $D$ values ranging from -1.5 to 1.5 as opposed to -4 to 4. }
	\label{fig:XMLSvsSPRCX_MM_Dzoom}
\end{figure}

Considering a more general problem, Figure~\ref{fig:XMLSvsSPRCX_MM_Dzoom} shows the local effectivity index $\efecl$ for the fourth mesh in a sequence of unstructured meshes, having a number of degrees of freedom (dof) similar to the mesh represented previously for load modes I and II. The XMLS recovered field presents higher overestimation of the error around the crack tip which corroborates the results found in previous load cases. 

\begin{figure}[!ht]
	\centering
	 \includegraphics{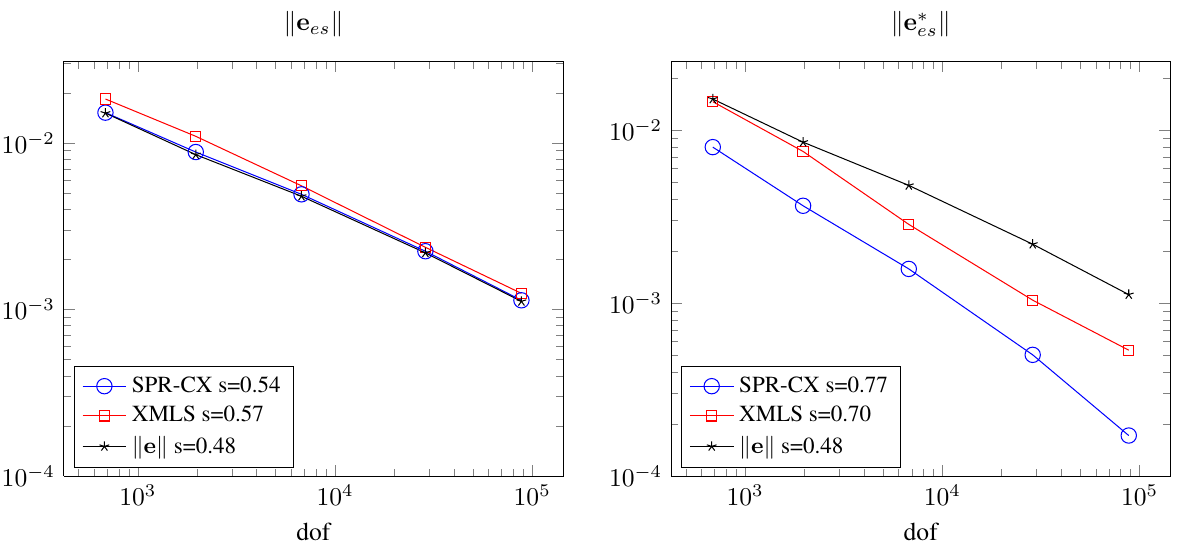}
	\caption{ Mixed mode, unstructured meshes: Convergence of the estimated error in the energy norm $\Vert \mathbf{e} _{es}\Vert$ (left). Convergence of the error for the recovered field $\Vert \mathbf{e} _{es}^*\Vert$ (right). The results are almost identical to the structured mesh case, except for the faster convergence obtained for SPR-CX in the unstructured compared to the structured case (0.77 versus 0.59), keeping in mind that optimal convergence rates are only formally obtained for structured meshes. }
	\label{fig:XMLSvsSPRCX_MM_enorm}
\end{figure}

\begin{figure}[!ht]
	\centering
	 \includegraphics{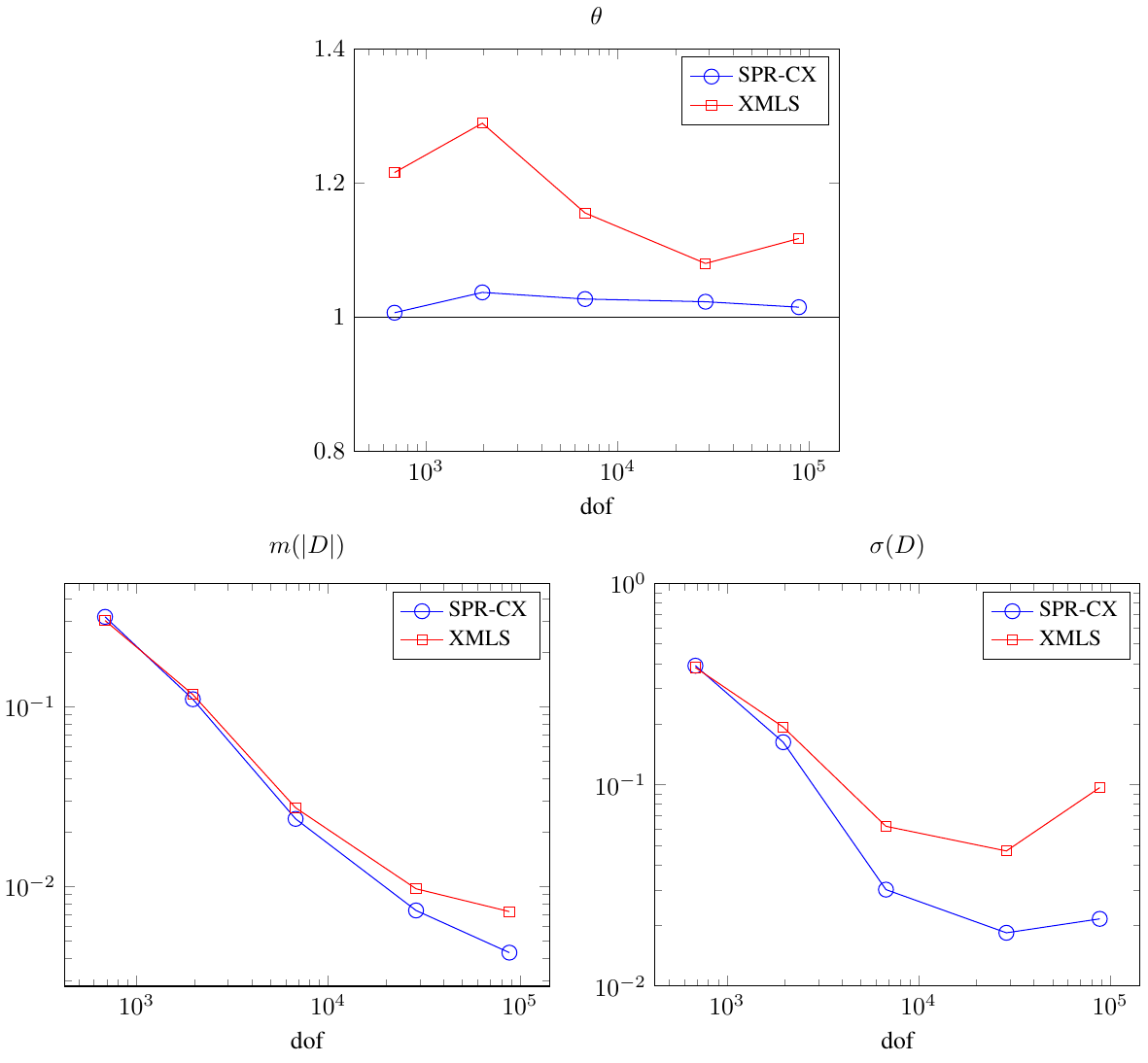}		
	\caption{Global indicators $\efec$, $\meanD$ and $\stdD$ for mixed mode and unstructured meshes. Notice that the XMLS performs more closely to SPR-CX using those indicators, for unstructured than for structured meshes, especially when measuring the mean value of the effectivities $\meanD$. }
	\label{fig:XMLSvsSPRCX_MM_Effec}
\end{figure}

Figures \ref{fig:XMLSvsSPRCX_MM_enorm} and \ref{fig:XMLSvsSPRCX_MM_Effec} represent the evolution of global parameters for unstructured meshes and mixed mode. Once more, the best results for the error estimates are obtained using the SPR-CX technique.

\section{Conclusions}

The aim of this paper was to assess the accuracy gains provided by 
\begin{enumerate}
\item Including relevant enrichment functions in the recovery process;
\item Enforcing statical admissibility of the recovered solution.	
\end{enumerate}

We focused on two  recovery-based error estimators already available for LEFM problems using the XFEM. The first technique called SPR-CX is an enhancement of the SPR-based error estimator presented by \cite{Rodenas2007a}, where the stress field is split into two parts (singular and smooth) and equilibrium equations are enforced locally on patches. The second technique is the XMLS proposed by \cite{Bordas2007, Bordas2008} which enriches the basis of MLS shape functions, and uses a diffraction criterion, in order to capture the discontinuity along the crack faces and the singularity at the crack tip.

To analyse the behaviour of the ZZ error estimator using both techniques and to assess the quality of the recovered stress field, we have evaluated the effectivity index considering problems with an exact solution. Convergence of the estimated error in the energy norm and other local error indicators are also evaluated. To analyse the influence of the special features introduced in the recovery process we have also considered two additional configurations: SPR-C and SPR-X.

The results indicate that both techniques, SPR-CX and XMLS, provide error estimates that converge to the exact value and can be considered as asymptotically exact. Both techniques could be effectively used to estimate the error in XFEM approximations, while other conventional recovery procedures which do not include the enrichment functions in the recovery process have proved not to converge to the exact error. This shows (albeit only numerically) the need for the use of extended recovery techniques for accuracy assessment in the XFEM context. 

Better results are systematically obtained when using the SPR-CX to recover the stress field, specially in the areas close to the singular point. For all the different test cases analysed, the XMLS produced higher values of the effectivity index in the enriched area, where the SPR-CX proved to be more accurate. This can be ascribed to the fact that the SPR-CX technique recovers the singular part of the solution using the known equilibrated exact expressions for the asymptotic fields around the crack tip and enforces the fulfilment of the equilibrium equations on patches. Further work currently in progress will include the development of an equilibrated XMLS formulation, which could provide a continuous and globally equilibrated recovered stress field which can then be used to obtain upper bounds of the error in energy norm. 

The aim of our project is to tackle practical engineering problems such as those presented in \cite{ Bordas2006, Wyart2007} where the accuracy of the stress intensity factor is the target. The superiority of SPR-CX may then be particularly relevant. To minimise the error on the stress intensity factors we will target this error directly, through goal-oriented error estimates. This will be reported in a forthcoming publication. However, although the SPR-CX results are superior in general, \red{an enhanced version of the XMLS technique presented in this paper where we enforce equilibrium conditions} could result useful to evaluate upper error bounds when considering goal-oriented error estimates, as it directly produces a continuous equilibrated  recovered stress field.
 
Our next step will be the comparison of available error estimators in three dimensional settings in terms of accuracy versus computational cost to minimise the error on the crack path and damage tolerance of the structure. 

\section{Acknowledgements}
 
\red{St\'{e}phane Bordas would like to thank the support of the Royal Academy of Engineering and of the Leverhulme Trust for his Senior Research Fellowship entitled ``Towards the next generation surgical simulators'' as well as the support of EPSRC under grant EP/G042705/1 Increased Reliability for Industrially Relevant Automatic Crack Growth Simulation with the eXtended Finite Element Method.\\
} 
\red{This work has been carried out within the framework of the research projects DPI 2007-66773-C02-01, DPI2010-20542 and DPI2010-20990 of the Ministerio de Ciencia e Innovación (Spain). Funding from {FEDER}, Universitat Politècnica de València and Generalitat Valenciana is also acknowledged.
}

\bibliographystyle{apa}
\bibliography{General_LATIN1}

\vspace{10pt}
\textbf{Corresponding author}\\
\red{Octavio Andrés González-Estrada can be contacted at: ocgones@upv.es
}

\end{document}